\documentclass{article}
\usepackage{PRIMEarxiv}

\usepackage{amsmath,amssymb}
\usepackage{bbm}
\usepackage{algorithm}
\usepackage{algpseudocode}
\usepackage{booktabs}

\usepackage{textcomp,marvosym}

\usepackage{cite}

\usepackage{nameref,hyperref}
\usepackage{comment}

\usepackage[table]{xcolor}

\usepackage{array}
\usepackage{dirtytalk}

\newcolumntype{+}{!{\vrule width 2pt}}

\newlength\savedwidth

\raggedright
\usepackage[aboveskip=1pt,labelfont=bf,labelsep=period,justification=raggedright,singlelinecheck=off]{caption}

\makeatletter
\renewcommand{\@biblabel}[1]{\quad#1.}
\makeatother

\usepackage{subcaption}
\usepackage{lastpage,fancyhdr,graphicx}
\usepackage{epstopdf}
\fancyheadoffset[L]{2.25in}
\fancyfootoffset[L]{2.25in}
\begin{document}
\vspace*{0.2in}

\begin{flushleft}
{\Large
\textbf\newline{Characterizing user archetypes and discussions on Scored.co} 
}
\newline
\\
Andrea Failla\textsuperscript{1,2*},
Salvatore Citraro\textsuperscript{2},
Giulio Rossetti\textsuperscript{2}, Francesco Cauteruccio\textsuperscript{3}
\\
\bigskip
\textbf{1} Department of Computer Science, University of Pisa, Pisa I56127, Italy
\\
\textbf{2} Institute of Information Science and Technologies “A. Faedo” (ISTI), National Research Council (CNR), Pisa I56127, Italy
\\
\textbf{3} DIEM, University of Salerno, Fisciano I84084, Italy
\\
\bigskip

%
%





*andrea.failla@phd.unipi.it

\end{flushleft}
\section*{Abstract}
In recent years, the proliferation of social platforms has drastically transformed the way individuals interact, organize, and share information. 
In this scenario, we experience an unprecedented increase in the scale and complexity of interactions and, at the same time, little to no research about some fringe social platforms. 
In this paper, we present a multi-dimensional framework for characterizing nodes and hyperedges in social hypernetworks, with a focus on the understudied alt-right platform Scored.co. 
Our approach integrates the possibility of studying higher-order interactions, thanks to the hypernetwork representation, and various node features such as user activity, sentiment, and toxicity, with the aim to define distinct user archetypes and understand their roles within the network. 
Utilizing a comprehensive dataset from Scored.co, we analyze the dynamics of these archetypes over time and explore their interactions and influence within the community. 
The framework's versatility allows for detailed analysis of both individual user behaviors and broader social structures. 
Our findings highlight the importance of higher-order interactions in understanding social dynamics, offering new insights into the roles and behaviors that emerge in complex online environments.


\section{Introduction}

In the last decades, the proliferation of social platforms and the opportunities tied to each of them have drastically transformed the way individuals interact, share information, and form communities. This unprecedented growth in the scale and complexity of online interactions is largely due to social networks like X (formerly known as Twitter), discussion fora like Reddit, and even more niche platforms, like the now-defunct Voat, known for their explorations of censorship-free models from the point of view of specific political groups~\cite{mekacher2022can}.
Hence, understanding user behavior in general, and especially within these platforms, has become a crucial area of research in computational social science~\cite{lazer2009computational,Cauteruccio*22}. 
By studying how users interact with each other and with others' content, researchers can uncover patterns of influence spread, opinion formation, and creation and dissolution of communities~\cite{failla2024describing}.
These insights are not only academically interesting but also have practical implications for designing better social platforms, enhancing user experience, and mitigating issues such as misinformation and online harassment~\cite{Record*18,Kou*18}.
Moreover, understanding these behaviors can aid in developing strategies for marketing, public health campaigns, and even political mobilization, making this research area highly relevant in multiple domains. Furthermore, all of these aspects are even more critical in the analysis of understudied platforms such as Scored.co~\cite{Patel*24}\footnote{\url{https://scored.co}}, which have not been extensively explored, and thus offering unique social dynamics that may differ significantly from mainstream platforms. Investigating these less-explored environments can uncover novel patterns of user behavior and mechanisms of information dissemination. Also, such platform 
often cater to niche communities or specialized interests, thus providing insights into subcultures and micro-level social dynamics that might not be straightforwardly visible on larger social platforms.

Understanding these dynamics becomes particularly important when addressing the general challenges inherent in studying complex social platforms.
Among all of these, a significant one is the identification of roles that users play within complex social structures. Indeed, role identification is crucial for various applications, including targeted information dissemination~\cite{ZhWuJi19} and community detection~\cite{Cauteruccio*22}. In traditional social network analysis, roles are often determined based on patterns of dyadic connectivity and interactions. 
Nevertheless, in a discussion on a social platform, two users can interact directly, e.g., by sharing comments, but also indirectly, i.e., through other users.
Hence, it is reasonable to assume that taking into account group dynamics and higher-order interactions could help in a more reliable identification of the roles played by users.
A promising approach to encapsulate group interactions is defined by the concept of a social hypernetwork \cite{Battiston*20,aksoy2020hypernetwork,failla2023attributed}, which allows for the modeling of relationships among groups of nodes, generally called hyperedges. Such a representation is particularly useful for studying social platforms from a higher-order point of view. 
Despite its importance and the existing methodologies designed for pairwise graphs, current literature about the problem of role identification lacks comprehensive approaches for social hypernetworks.

To address this gap, in this work, we formally introduce an approach for defining higher-order roles in a social hypernetwork and characterize higher-order interactions.
We define the concept of an archetype, a general characterization of higher-order roles that serves as a ``template'' for identifying nodes representative of specific roles within the context of hyperedges. These archetypes provide a structured way to categorize and understand the diverse roles users play in group interactions. Then, we propose a general hyperedge characterization function, which allows one to characterize a hyperedge w.r.t. features of the nodes contained in it. To validate the effectiveness of our proposed framework, we conduct an exhaustive experimental campaign on the understudied social platform Scored.co, which offers a rich, novel dataset for analyzing user interactions in a hypernetwork context, making it an ideal testbed for our framework. Our experiments demonstrate the utility of our approach in characterizing user roles within social hypernetworks, and allow us to derive several insights and implications enabled by our framework.

Summarizing, the main contributions of this work are as follows:

\begin{itemize}
    \item We define the concept of archetype, as a high-level description of user roles acting as a ``template'' characterizing node representatives. 
    We also propose a way of quantifying how \say{distant} a node is from its archetypal representation; 
    \item We introduce an approach to characterize higher-order entities, such as hyperedges, taking into account exhibited features and their higher-order dynamics;
    \item We apply our framework to a newly collected dataset of content and interactions from Scored.co, which we also release (anonymized)~\cite{scoredHD}.
    Notably, this is also the first work studying the structural properties of Scored.co.
\end{itemize}

 We believe that our contribution could be pivotal w.r.t. different aspects, as we will thoroughly illustrate in the rest of the paper. Indeed, understanding user roles within complex social structures is a key task in computational social science, especially on platforms where interactions extend beyond dyadic relationships. By defining archetypes as role-based templates within a hypernetwork, our framework provides a better understanding of user behaviors in settings where interactions involve multiple participants simultaneously. Strictly related is the analysis of group-based dynamics that are not visible through dyadic connections alone, which is made possible by our proposed approach for higher-order entity characterization and exhibits a practical utility in understanding complex interactions within both social and heterogeneous systems. 

The outline of this paper is as follows: in the~\nameref{sec:related-literature} section, we provide a detailed overview of related literature. In the~\nameref{sec:framework} section, we provide the formalization of our proposed approach by defining the concept of archetype first, and the hyperedge characterization function after. In the~\nameref{sec:experiments} section, we illustrate the experimental campaign we carried out to evaluate the effectiveness of our framework. Afterward, in the~\nameref{sec:discussion} section we propose a discussion regarding our approach and we highlight some important implications. Finally, in the~\nameref{sec:conclusion} section, we draw our conclusions and delineate some future works.

\section{Related Literature}
\label{sec:related-literature}


The identification of roles within social networks stands as a fundamental aspect for understanding the dynamics and function of these complex systems~\cite{SeStLe16}. In traditional social network analysis, roles are often determined based on patterns of connectivity that provide insights into the influence, responsibilities, and grouping of individuals within the network~\cite{AlGaradi*18}. This analysis is crucial for applications ranging from the characterization of polarization~\cite{ReZaSo19} and user behaviors~\cite{CaKo23} to the analysis of biological networks~\cite{WaLuYu14} and animal populations~\cite{Bajardi*11}. In different cases, role identification is also framed within the task of discovering influential users in a social network. In light of this, we refer the interested reader to the comprehensive survey proposed in~\cite{AlGaradi*18}.

Nevertheless, extending role identification into hypernetwork scenarios poses many challenges. Unlike traditional networks, hypernetworks consist of interactions, represented by hyperedges, that can encompass multiple nodes simultaneously. 
Higher-order interaction models can better represent complex real-world systems ~\cite{torres2021and,Battiston*20}; however, the complexity of higher-order interactions can introduce difficulties in accurately defining and identifying roles, as traditional centrality metrics and methodologies (e.g., community detection algorithms) often fall short in capturing the multidimensional nature of hyperedges~\cite{PaPeVa17,BoHoJo04,SeStLe16}.
Despite the relevance of such a task, to the best of our knowledge, there are no existing papers that explicitly discuss role identification within the context of hypernetworks.
Given this gap in the literature, the rest of the section will concentrate on providing an overview of the methodologies and findings from the conventional settings in dyadic interactions, which may serve as a baseline for future exploration into hypernetwork-specific role identification. 


The general task of role identification consists in capturing the multifaceted characteristics of a node's role in a pairwise network. The work in~\cite{Huang*14} identifies roles in undirected, unweighted networks by evaluating node importance using correlated indicators like degree and centrality measures. 
A node's role is then determined by comparing its indicator relationships to statistical correlations within the overall network. While this work shares the task of role identification with ours, the process does not involve group-based indicators leveraging the concept of ``groups" such as communities or subgraphs.
The work in~\cite{BhCoMu11} frames the node role identification tasks as a node classification approach based on random walks.
Similarly, framed within a classification tasks, the authors in~\cite{BuGo14} focus on the so-called ``answer-person'' role on Reddit, characterized by users who predominantly respond to questions posed by others, with minimal engagement in broader discussions.
Both the works in~\cite{BhCoMu11} and \cite{BuGo14} share a similar aim with ours. Nevertheless, they rely on supervised approaches on pairwise graphs. Also, the methods might not be straightforwardly generalized to higher-order interactions. 

The approach proposed in~\cite{BrKr11} focuses on the identification of node roles in dynamic social networks as a sequence of different types of activities, by leveraging pattern subgraphs and sequence diagrams.
Doing so, the authors are able to capture roles such as the ``gossipmongers", i.e., users who replicate every received message at least three times. The work in~\cite{BrKr11} can be considered as orthogonal to ours, as an hypergraph mining approach could also be applied in our context to enhance role identification.
The authors in~\cite{HaRi21} outline a process for identifying user roles in Enterprise Social Networks (ESNs), using a mix of design science research and data mining, involving data collection, preparation, and evaluation, with user roles identified through statistical analysis, including PCA and clustering.
The work in~\cite{HaRi21} shares a similar focus with ours, although using a different methodology.
The identification of emergent leadership roles is addressed in~\cite{Temdee*06}, where the authors examine collaboration patterns between teams and propose the so-called ``leadership index", a combination of centrality measures including closeness and betweenness. 
In~\cite{ZhWuJi19}, the authors propose a mixed-methods methodology for user role identification, focusing on dynamic user profiling. 
Several types of special users are defined and identified to support information dissemination.
The work in~\cite{ZhWuJi19} proposes an approach based on computational methods and questionnaire-based evaluation to quantitatively describe user features. Such work parallels ours in the sense that it considers the network's features, too.
In~\cite{Cauteruccio*22}, the authors study community and user stereotypes in Reddit. Here, with the term stereotype, the authors refer to the tendency to classify people into groups and to associate each group with a general idea or a label. The authors propose a rule-based approach based on different quantitative views of the data, and define author stereotypes on the basis of two orthogonal taxonomies, namely, the number of posts, and the number of comments of an author. While the work in~\cite{Cauteruccio*22} and ours share some similarities, the employed methodologies are substantially different. The former defines stereotypes based on a single quantitative measure called score, which is inherent to the social platform itself. Instead, our approach proposes a general framework in which the characterization of users and the identification of their role are both defined by their feature-rich, higher-order surroundings. 
In~\cite{Kou*18}, the authors identify five distinct social roles in a specific community on Reddit. Among them, we cite the ``knowledge broker'', i.e., a member who introduces knowledge to the community by sharing links, and the ``translator'', i.e., a member who contributes academic knowledge. While the contribution in~\cite{Kou*18} is notable and similar to ours for considering various aspects of user characterization, these roles are specific and applicable only to the analyzed community.

Finally, it is worth mentioning embedding-based methodologies~\cite{Zhou*22}, which could potentially be adapted to hypernetwork scenarios. For instance, in~\cite{RoAlSa21}, the authors introduce a technique that incorporates node attributes to generate embeddings that capture the similarities based on neighborhood structures. Instead, in~\cite{Dehghan*23}, the authors leverage both node and structural embeddings to detect nodes impersonating social bots. Indeed, node embeddings based on structural properties and exhibited attributes attracted substantial interest in the last years~\cite{Junchen*21}. While these methods use embedding techniques that effectively summarize and exploit the structural information within traditional networks, their application to hypernetworks might not be straightforward.
Our approach diverges from embedding-based methods in the sense that it is specifically tailored to the unique features of nodes within hypernetworks, such as their exhibited features w.r.t. the hyperedges containing them.
By focusing on this, without relying on an embedding, our methodology seeks to offer a more direct and fitting analysis of node roles.


 We conclude this bird's-eye view of the related literature by focusing on the correlated aspect of understudied7fringe platforms. Intuitively, the reasoning provided in the above discussion of related work also holds for these platforms. 
In fact, there is a moderate number of works on understudied platforms, and generally, these are limited to presenting a dataset collection about the considered platform~\cite{Patel*24,FaRo24,QuBo24,MeFaBa24}. 
For instance, a large-scale dataset of the social platform Scored.co has been presented in~\cite{Patel*24}, in which the authors studied aspects such as posting activity and user characterization, as well as the phenomenon of user migration from other platforms. A recent social platform, called Bluesky, has been studied in~\cite{FaRo24, QuBo24}. In the former work, the authors presented a comprehensive study of the social structure of the platform, as well as posting activity and content analysis. Also, the complete post history of over 4M users has been released. Similarly, the latter work studied users' political leaning and ideological polarization on BlueSky while also presenting a characterization of the network topology over time. Finally, in~\cite{MeFaBa24}, the authors released a dataset targeting the Indian microblogging platform Koo. The dataset consists of more than 72M posts and 75M comments, with related features such as shares and likes. Also, a thorough overview of the platform is presented, consisting of a discussion of the news ecosystem on the platform, hashtag usage, and user engagement.

\section{Materials and methods}
\label{sec:framework}

In this section, we introduce our framework to characterize nodes and hyperedges in a social hypernetwork, as well as the definition of roles, called archetypes, in such a network. We start by providing some background, which is useful to understand the frame of our context. Then, we introduce the definition of archetypes and our method to calculate them. Finally, we detail the proposed framework to characterize nodes and hyperedges.

\subsection{Background}
\begin{figure}
    \centering
    \includegraphics[width=0.5\linewidth]{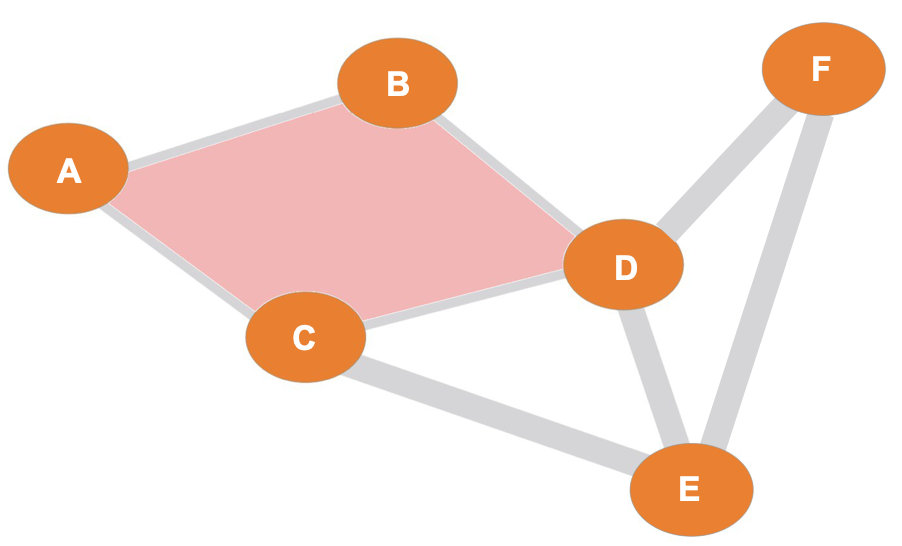}
    \caption{A toy hypergraph. Nodes are labelled with capital letters. 
    Nodes A, B, C and D are connected by a hyperedge of size 4.
    Nodes C and E, D and E, D and F, and E and F are pairwise connected.
    }
    \label{fig:hg}
\end{figure}
A hypergraph $H=(V,E)$, $V = \{v_1, \dots, v_n\}$ is the set of nodes, $E = \{e_1, \dots, e_m\}$ is the set of hyperedges, where $e_j \subseteq V$, for $j = 1,\dots,m$. 
A visual representation of a hypergraph is depicted in Fig.~\ref{fig:hg}.
The order of $H$ is $n = |V|$, while the size of $H$ is $m = |E|$. We denote with $E_{v_i}$ the set of hyperedges containing $v_i$, that is $E_{v_i} = \{e_j \in E : v_i \in e_j\}$. Given a hyperedge $e_j \in E$, its size is the number of nodes belonging to it, that is $|e_j|$. The degree of a node $v_i$, denoted as $deg(v_i)$, is the number of neighbors of $v_i$; a node $v_k \in V$ is the neighbor of node $v_i$ if and only if there exists at least one hyperedge $e_j$ which $v_i$ and $v_k$ both belong to. The hyperdegree of a node $v_i$, denoted as $hdeg(v_i)$, is the number of hyperedges to which $v_i$ belongs. A graphical depiction of a hypergraph is given in Fig.~\ref{fig:hg}.

We use a hypergraph $H$ to model a social hypernetwork. Here, nodes are users of the social hypernetwork, while a hyperedge represents a discussion between users. In what follows, we will refer to the social hypernetwork simply as hypernetwork.

Moreover, features can be associated with users, and these are generally based on the content the users interact with and the discussions they participate in. To formalize them, we employ a set of features $F = \{f_1,\dots,f_l\}$ characterizing the nodes in $H$. Given a node $v_i \in V$, $F_{i} = \{f_{1_i},\dots,f_{l_i}\}$ represents the values of each feature of $v_i$, that is $f_{k_i}$ indicates the value of the $k$-th feature of $v_i$. We assume features can be either numerical or categorical and that numerical ones are always normalized. Also, with $H_F$, we indicate the hypergraph equipped with the set of node features $F$.

\subsection{Characterizing Nodes via Archetypes}
\label{sec:archetype}

The first aim of our approach is the definition of archetypes. An archetype serves as a ``template'' to represent nodes characterized by a certain subset of features from the set $F$. 
The concept of an archetype is pivotal in characterizing higher-order node roles based on subsets of hyperedges. Often, the study of a representation of nodes in a social network involves the definition of a taxonomy or a more thorough analysis of the behavior of such nodes~\cite{Cauteruccio*22}. Instead, our definition of archetype enables a more general characterization of possible behavioral dynamics occurring among nodes, and does not rely on a single attribute.

Let $F_A = \{f_1, \ldots, f_p\} \subseteq F$ be a subset of features. Then, an archetype $A$ is defined as a tuple of values based on the features indicated by $F_A$. Formally, let $A = \langle a_1, a_2, \ldots, a_p \rangle$, where $a_j$ is the value of the $j$-th feature from $F_A$ for the archetype $A$. Then, each $a_k$, for $k = 1,\ldots, p$, represents a particular value of the $k$-th feature from $F_A$ that characterizes the nodes associated with this archetype. As an example, let $F_A = \{f_{\texttt{sent}}, f_{\texttt{tox}}\}$, where $f_{\texttt{sent}}$ (resp., $f_{\texttt{tox}}$) is the value of a quantitative feature indicating the average sentiment value (resp., average toxicity value) expressed by a user. Such features are extensively used in various data science-based studies and can be easily computed via classical methods such as VADER~\cite{HuEr14} and Detoxify~\cite{Detoxify}. Suppose $f_{\texttt{sent}} \in [-1,1]$, where $-1$ (resp., $1$) indicates a mostly negative (resp., mostly positive) average sentiment, and $f_{\texttt{tox}} \in [0,1]$, with higher values indicating a high degree of toxicity. Then, various archetypes could be defined based on these features. For instance, we could define the archetype $A = \langle -1, 0.5 \rangle$ that describes a template for users that are extremely negative and are toxic on average, and we could name this archetype as \emph{Cinically Commenter}; instead, the archetype $A = \langle 1, 0.5 \rangle$ would be a template for users that are extremely positive and exhibit moderate toxicity, and we could name it as \emph{Overzealous User}. Also, it is noteworthy that, for the sake of presentation, here the definition of archetype is only given w.r.t. nodes. Nevertheless, in case of features being also available for hyperedges, then the same definition applies, and archetypes for the latter can be defined.

Essentially, an archetype $A$ can be viewed as a prototypical example of nodes that share similar feature values. To simplify the analysis and categorization of archetypes, instead of taking into account the feature values directly, we can map them into categorical values, indicating the state of the feature for the particular archetype. This approach involves setting a series of thresholds $T = \langle t_1, \dots, t_p \rangle$, where the value $t_j$ has the same domain of the feature $f_j$ in $F$, a set of labels $L$ and a labeling function $\iota$. Then, given an archetype $A = \langle a_1, \dots, a_p \rangle$, by applying to each feature $a_j$ the corresponding threshold $t_j$ from $T$ via the labeling function $\iota$, we derive the archetype $A_T = \langle a^t_1, \dots, a^t_2 \rangle$, where $\iota(a_j,t_j) = a^t_j$ is a label in $L$. Practically speaking, $A_T$ represents the same archetype as $A$, but its representation is now built over a specific set of labels. Let us take the aforementioned archetype $A = \langle 1, 0.5\rangle$, defined over the set of features $F_A=\{f_{\texttt{sent}}, f_\texttt{tox}\}$, and representing Overzealous Users. Suppose we set $T = \langle 0, 0.75 \rangle$, and we set $L = \{\texttt{low}, \texttt{high}\}$. Also, suppose we define the labeling function $\iota$ as

\[
\iota(a_j,t_j) = \begin{cases}
\texttt{low} & \text{if} \ a_j \leq t_j \\
\texttt{high} & \text{otherwise}
\end{cases} 
\]

\noindent Therefore, in this example, we derive $A_T = \langle \texttt{high}, \texttt{low} \rangle$, which can be seamlessly interpreted as representing users exhibiting a high sentiment value and a low toxicity value. Note how the labeling depends on the thresholds $T$, thus allowing for flexibility that can accommodate different contexts of analysis.

Finally, we are now able to state when a node is represented by a given archetype. Given an archetype $A=\langle a_i, a_2, \dots, a_p \rangle$, based on a subset of features $F_A = \{f_1, \ldots, f_p\} \subseteq F$, we want to effectively understand what nodes are represented by it. Let $v_i \in V$ be a node, and let $F_{A}(v_i) = \langle f_{i_1}, f_{i_2}, \dots, f_{i_p} \rangle$ be the feature vector of node $v_i$ according to the features selected in $F_{A}$. The node $v_i$ can be considered as represented by the archetype $A$ if $F_{A}(v_i)$ is sufficiently close to $A$, according to a predefined distance metric $d$. We express it as $ d(F_{A}(v_i), A) \leq \epsilon $ where $\epsilon$ is a small positive threshold value that determines the acceptable distance between the node's feature vector and the archetype. Note that both $A$ and $F_{A}(v_i)$ can be considered as vectors of the same length $p$; thus, classical distance metrics, such as the cosine similarity~\cite{Singhal01}, can be used. Furthermore, as we will see in the experiments, in the simplest case also a feature-wise comparison can be used to assess when a node can be considered as represented by a given archetype.

\subsubsection{A Characterization for Archetypes}
\label{sec:archetype-characterization}

To comprehensively understand and categorize user archetypes, we decide to observe them through three different expressions, namely, \textit{(i)} emotional, \textit{(ii)} psycho-emotional, and \textit{(iii)} moral expressions. By leveraging well-established psychological theories and lexicons, we aim to create detailed profiles that reflect the peculiar ways in which users interact and participate within the social platform. In what follows, we describe in detail these three expressions, which we will subsequently use in our experiments.

\paragraph*{\textbf{Emotional Profiles}} We aim to characterize archetypes based on the emotions they express.
To do so, we refer to the psychological theory of emotions~\cite{plutchik1980general} developed in 1980 by the American psychologist Robert Plutchik.
This theory identifies eight basic emotions --- joy, trust, fear, surprise, sadness, anticipation, anger, and disgust --- and claims that all other emotions derive from a mixture of these primary ones.
To quantify feelings expressed by Scored users, we leverage the NRCLexicon (National Research Council Lexicon)~\cite{mohammad2013crowdsourcing}, a resource containing over 14,000 English words and their associated emotional ratings according to Plutchik’s theory.
This dictionary was further expanded by the National Research Council of Canada to include WordNet synonyms, reaching over 27,000 terms\footnote{\url{https://github.com/metalcorebear/NRCLex}}. 
For each of the users' texts, we compute emotion scores, and normalize them in [0, 1]. 
In this context, 0 indicates texts that do not elicit any emotion, while 1 signifies texts that strongly evoke the specified emotion.

\paragraph*{\textbf{Psycho-emotional Profiles}} We also characterize how user archetypes relate to their surrounding social environments.
To do so, we refer to the PAD model (Pleasure, Arousal, Dominance) introduced by Mehrabian and Russell in 1974~\cite{mehrabian1974approach}.
According to the PAD model, three dimensions characterize the perception an individual has of the environment in which she finds herself.
Pleasure (sometimes referred to as \textit{valence}) concerns whether an individual perceives the environment as enjoyable or not. 
Arousal measures how stimulating the environment is for the individual. 
Dominance indicates whether the individual feels in control of the environment.
To operationalize these dimensions, we leverage the VAD Lexicon (Valence, Arousal, Dominance)~\cite{mohammad2018obtaining}. 
This resource contains over 20,000 English
terms, along with their associated valence/pleasure, arousal, and dominance values. 
For each of the three dimensions, we associate each text with the total score of its words.
Then, we normalize results in [0,1], where 0 implies an absence of the corresponding dimension, and 1 implies the strong presence thereof.

\paragraph*{\textbf{Moral Profiles}}
We aim to characterize archetypes based on the moral dimensions that emerge from the content they produce.
We rely on the Moral Foundations Theory, a psychological framing rooted in cultural anthropology that postulates the existence of five universal moral dimensions~\cite{haidt2004intuitive}: \textit{authority/subversion}, \textit{care/harm}, \textit{fairness/cheating}, \textit{loyalty/betrayal}, and \textit{sanctity/degradation}.
Each dimension is composed of a virtue (e.g., loyalty), and a corresponding vice (e.g., betrayal).
Virtues can be understood as follows, while vices can be considered their opposites.
The concept of authority can be defined in relation to specific traits, such as deference to higher authorities, in order to maintain group cohesion. 
Similarly, the concept of care can be understood in terms of nurturing and protection. 
Fairness can be conceptualized in terms of equal treatment and reward. 
Loyalty can be understood in relation to the prioritization of one's group and alliances. 
Finally, sanctity can be defined in terms of the maintenance of the sacredness of the body and the avoidance of moral contamination.
We operationalize this framework via the eMFD (Extended Moral Foundations Dictionary), a lexicon containing more than 3,000 words~\cite{hopp2021extended}.
Each word has an associated score in [-1, 1] for each foundation, ranging from strong vice outage (-1) to strong virtue outage (1).

\subsection{Analyzing Higher-Order Entities}
\label{sub:hyperedge-characterization-function}

While archetypes are a particular yet effective way to analyze entities within the social platform, they mainly focus on the node features, whereas interactions are not taken into account.
Hence, we define here a general characterization function to characterize nodes and hyperedges according to their exhibited features and higher-order dynamics. 
Without loss of generality, we propose the definition of such a function w.r.t. hyperedges. The same can be also applied to nodes.

Let $H_F = (V,E)$ be a hypergraph equipped with the node features set $F$. We denote with $\omega$ a function which we call the \textit{hyperedge characterization function}. The need for a function such as $\omega$ addresses the challenge of characterizing a hyperedge w.r.t. the nodes contained in it.
Let us recall that our approach deals with analyzing higher-order entities. To do so, we exploit the representation of node relationships via hyperedges. While it effectively captures the higher-order structural interactions between nodes, it might not be sufficient in acquiring insights into the semantics of such interactions. Therefore, we focus on the latter aspect through a characterization of hyperedges that is based not only on the contained nodes but also on their features. Formally, the domain of our hyperedge characterization function is $E$. $\omega$ takes in input a hyperedge $e \in E$ and returns a value $\omega(e_j)$, which we call its characteristic value. Such value depends on the actual implementation $\omega$: in fact, $\omega$ is general, and different approaches can be exploited to accommodate the hyperedge characterization. Furthermore, to address the aforementioned challenge, there are cases in which $\omega$ should be defined to consider the values of each node's features contained in the considered hyperedge. Given a feature of interest $f_k$, we write $\omega^k$ to denote that the actual implementation of $\omega$ considers the feature $f_k$.

In the following, we propose various specializations of $\omega$ and a brief rationale for each of them. Some of these specializations are used in the~\nameref{sec:experiments} section. We separate them in three families, namely, \textit{(i)} numerical-only definitions, \textit{(ii)} categorical-only, \textit{(iii)} structural-based ones. In describing each of them, we assume we are interested in characterizing a hyperedge $e \in E$. Also, there are different specializations that are intended to be exploited when the analysis we are carrying out is feature-oriented. Therefore, in these cases, we assume having a feature $f_k$ of interest.

\paragraph*{Numerical-only specializations} The following specializations are intended to be used when numerical features are considered within the investigation. Therefore, here we assume the feature $f_k$ of interest is numerical. Some specializations are:

\begin{itemize}
    \item \textit{Statistics Descriptors}: common statistics descriptors such as mean, median, mode, variance, and standard deviation of the feature $f_i$ among the nodes in $e$ can be easily computed. For instance, the mean would be simply defined and denoted as $\omega_{\text{avg}} = \frac{\sum_{u \in e} f_{i_u}}{|e|}$.
    \item \textit{MAD}: calculates the Mean Absolute Deviation of the feature $f_i$ among the nodes in $e$, that is $\omega_{\text{mad}}(e) = \sum_{u \in e} \frac{|f_{i_u} - \omega_{\text{avg}}|}{|e|}$.
    \item \textit{Gini Coefficient}: employs the Gini Coefficient to compute the dispersion of the values of $f_i$ among the nodes in $e$, that is $\omega_{\text{Gini}}(e) = \sum_{u \in e}\sum_{v \in e, u \neq v} \frac{|f_{i_u} - f_{i_v}|}{2 \overline{f_i} |e|^2}$. 
\end{itemize}

\paragraph*{Categorical-only definitions} Different from the previous ones, the following specializations are intended to be used when categorical features are considered. Therefore, here we assume the feature $f_k$ of interest is categorical. Some specializations are as follows:

\begin{itemize}
    \item \textit{Entropy}: measures the cohesion of the values of $f_i$ among the nodes in $e$. We define and denote it as $\omega_{\text{entr}}(e) = -\sum_{u \in e}r_{i_u}log(r_{i_u})$, where $r_{i_u}$ denotes the proportion of the value of feature $f_{i_u}$ over all nodes in $e$.
    \item \textit{Gini Impurity}: measures the probability of incorrectly classifying a randomly chosen element in the dataset if it were randomly labeled according to the distribution of all categories in the dataset. Here, we employ it on the feature $f_i$, thus we define and denote it as $\omega_{\text{GiniImp}}(e) = 1 - \sum_{u \in e} r_{i_u}^2$, where $r_{i_u}$ denotes the proportion of the value of feature $f_{i_u}$ over all nodes in $e$.
\end{itemize}

\paragraph*{Structural-based specializations}

The structural-based specializations focus on characterizing a hyperedge based on its exhibited structural properties rather than only focusing on the features of the contained nodes. Some proposed specializations are the following:

\begin{itemize}
    \item \textit{Hyperedge Size}: the measure is simply given by the size of the hyperedge $e$, that is $\omega_{\text{size}}(e)=|e|$, regardless of the feature $f_i$.
    
    \item \textit{Purity}: Here, the measure indicates the homogeneity of the values of $f_i$ in $e$, that is $\omega_{\text{purity}}(e)=\frac{max({count(f_{i_u}) : u \in e})}{|e|}$.



    \item \textit{Cohesion}: the characterization of the hyperedge $e$ is given by the mean pairwise similarity of the value of the feature $f_i$ for each pair of node $u, v \in e$. Formally, given a similarity function $sim(\cdot,\cdot)$ defined for two features, we define $\omega_{\text{cohes}}(e) = \frac{2}{|e|(|e|-1)}\sum_{u \in e}\sum_{v \in e, v \neq u} sim(f_{i_u}, f_{i_v})$.

    \item \textit{Interaction potential}: the hyperedge $e$ is characterized by its capacity to create external connections, i.e., connections that span outside the hyperedge. It measures the extent to which the hyperedge facilitates interactions beyond its mere composition. Formally, $\omega_{\text{intpot}}(e) = \frac{|\{ v \in V \setminus e : \exists u \in e, v \sim u \}|}{|e|}$, where $v \sim u$ indicates that $v$ and $u$ are neighbors within a hyperedge different from $e$. The denominator can be $|e|$, i.e., only the nodes in $e$ are taken into account, or $|V \setminus e|$, i.e., $\omega_{\text{intpot}}$ is sensitive to the order of the hypergraph.

\end{itemize}

\section{Experiments}
\label{sec:experiments}

In what follows, we first provide a characterization of the understudied dataset regarding the social platform Scored.co. We study archetypes based on platform-specific and linguistic features, and we illustrate how our proposed framework allows for a comprehensive analysis of them. Then, we focus on the constructed hypernetworks, and we provide a characterization of discussions, that is, hyperedges, through our proposed hyperedge characterization function.

\subsection{Dataset}

We focus on the \textit{Scored.co} social platform, which hosts a network of communities that users can create and join. The platform shares some similarities with the more well-known platform Reddit\footnote{\url{https://reddit.com}}. Users can join communities related to various topics, from politics to memes, and upvote content they like, and downvote content they dislike. The number of upvotes minus the downvotes constitutes a post's score, i.e., a value that indicates the content's usefulness and/or relevance to the discussion.

Interestingly, \textit{Scored.co} includes many far-right/alt-right communities after the 2020 massive Reddit bans~\cite{cima2024great}. Some of these maintained their names and key figures, e.g., \textsf{c/TheDonald}, \textsf{c/GreatAwakening}, ultimately acting as a continuation of the original ones.
\textit{Scored.co} is an emerging yet understudied platform, potentially hosting dangerous content. Despite its growing user base, it has received relatively little academic attention compared to other social media platforms, as we pointed out in the \nameref{sec:related-literature} section. The loosely moderated nature of some communities can lead to the spread of misinformation, hate speech, and radical ideologies. 
This makes \textit{Scored.co} a platform of interest for researchers studying online extremism, digital sociology, and the impacts of social media on public discourse.

To build the dataset for our experiments, we collect data from the \textit{Scored.co} platform via the official API\footnote{\url{https://help.scored.co/knowledge-base/getting-started-with-the-api/}}. To retrieve as much content as possible, we use a breadth-first search technique by first gathering all posts and comments authored by \texttt{C}, one of the platform's admins, which is one of the most active users on the platform. Then, the content is parsed to retrieve other users and their related content. We repeat this procedure until no new user is found. Eventually, we find 207,554 unique users (referenced by their username), 4,398,074 posts, and 36,978,685 comments. Our data collection process started on June 9th, 2024, and ended on June 22nd, 2024. The final dataset comprises higher-order interactions on Scored.co starting from the platform's launch on October 15th, 2019, and ending on June 1st, 2024.

The dataset is released on Zenodo~\cite{scoredHD}, anonymized and providing information for (i) higher-order interactions extracted from discussion threads and (ii) user profiles as sets of features characterizing each user at every time stamp.
A Python data collection script is also enclosed in the repository.


%

\subsection{Higher-order Interactions and Analysis}

Given the previously described dataset, we are now able to construct the hypernetworks which constitutes the basis of the experiments.
We start by constructing a hypernetwork ${\cal H} = (V,E)$, where $V$ denotes the set of Scored.co users who participated in at least a discussion, while $E$ denotes the set of hyperedges encoding discussion threads. That is, a user $v$ appears in $V$ if they participate in at least one discussion. Also, we discard all hyperedges of size 2 due to the fact that we are interested in studying higher-order interactions. This means that, for all $e \in E$, $|e| > 2$.
Since discussion topics on Scored are varied, ranging from politics to sports, memes, and more, we decided to center our analysis on discussions taking place in \textsf{c/TheDonald} during 2023. 
This community is a controversial and highly active online group that originally formed on Reddit before migrating to platforms like Scored.co~\cite{cima2024great}. We chose it because it is the oldest and most active community on the platform, covering 74\% of all collected posts and comments. 
This allows us to maintain topic continuity without sacrificing large portions of the data.
Note that ${\cal H}$ is a hypernetwork that represents the whole dataset. To enable analyses on the time dimension, we construct a hypergraph ${\cal H}_t = (V_t, E_t)$, for each month $t = 1, \dots, 12$, containing users and discussions relative to month $t$.
In Table~\ref{tab:stats}, we report summary statistics of all the constructed hypernetworks. The table reports basic statics, such as nodes and hyperedges number $|V|$ and $|E|$, respectively, the largest hyperedge size $max_{|e|}$, the average hyperdegree $\overline{hdeg}$, and the average degree $\overline{deg}$. Also, for each pair of adjacent timestamps $1 \leq t, t_{+1} < 12$, we report the Jaccard similarity index between node sets. The last row of the table reports the same information for the hypernetwork ${\cal H}$.

\begin{table}[!ht]
\centering
\caption{Statistics of the constructed hypergraphs. From left to right: number of nodes, number of hyperedges, largest hyperedge size, average hyperdegree, average number of neighbors, and Jaccard similarity index between adjacent timestamps.
The bottom row refers to the aggregated hypergraph $\mathcal{H}$.}
\label{tab:stats}
\begin{tabular}{lcccccc}
        \toprule
        $t$ & $|V|$ & $|E|$ & $max_{|e|}$ & $\overline{hdeg}$ & $\overline{deg}$ & $Jaccard_{t, t+1}$ \\
        \midrule
        1 & 10889 & 28280 & 293 & 19.59 & 394.17 & 0.63 \\
        2 & 10018 & 24234 & 230 & 18.24 & 348.39 & 0.63 \\
        3 & 10245 & 26393 & 411 & 19.18 & 382.09 & 0.63 \\
        4 & 10203 & 25406 & 428 & 18.72 & 375.51 & 0.64 \\
        5 & 9922 & 25681 & 368 & 19.43 & 391.11 & 0.64 \\
        6 & 9594 & 24558 & 357 & 19.13 & 359.61 & 0.63 \\
        7 & 9400 & 23869 & 225 & 19.00 & 335.22 & 0.62 \\
        8 & 9449 & 24542 & 380 & 19.30 & 358.27 & 0.63 \\
        9 & 9022 & 24132 & 454 & 19.90 & 364.10 & 0.63 \\
        10 & 9179 & 29632 & 241 & 23.39 & 376.16 & 0.64 \\
        11 & 8865 & 24962 & 258 & 20.77 & 351.09 & 0.65 \\
        12 & 8402 & 22711 & 216 & 20.34 & 333.75 & - \\ \midrule
        \textit{all} & 20937 & 321860 & 454 & 112.8 & 1090.89 & - \\
        \bottomrule
\end{tabular}
\end{table}

\begin{figure}
    \centering
    \includegraphics[width=\linewidth]{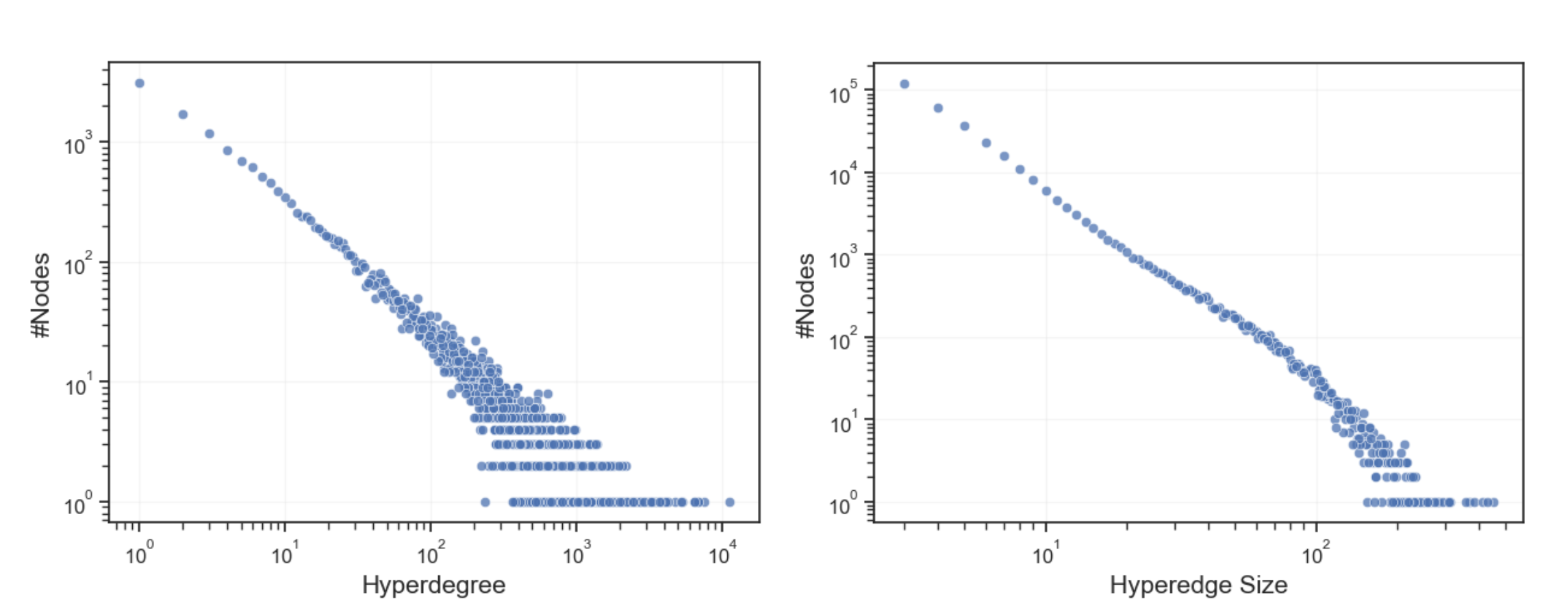}
    \caption{Hyperdegree (left) and hyperedge size (right)  distributions for the aggregated hypergraph}
    \label{fig:hedist}
\end{figure}

From the analysis of this table, we can observe different insights. First off, the aggregated hypernetwork $\mathcal{H}$ includes interactions among roughly 20\% of all discovered users. Indeed, on average, users participate in 113 discussions (hyperedges) throughout the year, and in 18-20 discussions each month. Scored.co users are exposed to $\sim 1K$ unique peers, with a monthly average of 300-400, thus showing locally dense patterns consistently over time. We can also observe how the largest hyperedge size exhibits some variations but remains in the same order throughout the whole period. This indicates the presence of a highly connected subgroup at certain timestamps. The average hyperdegree $\overline{hdeg}$ remains relatively stable across timestamps, indicating that on average, each node is part of $\sim 20$ hyperedges. The average degree $\overline{deg}$ remains fairly consistent, too. To complement these statistics, we try observing with a more detailed look the distribution of hyperdegree and hyperedge size, which are depicted in Figures~\ref{fig:hedist} (a) and (b), respectively. From the analysis of these figures, we can see that both show the typical power-law behavior observed in many real-world hypernetworks~\cite{patania2017shape,}. We observe how a few users participate in a large number of discussions, while most users participate in just a few. The hyperedge size distribution (Fig.~\ref{fig:hedist} (b)) displays a similar shape, with mostly small discussions and a few large ones. We also note how these patterns are consistent over time, with 44\% interactions of at least size 5, 15\% interactions of at least size 10, and 2\% interactions of at least size 50. Finally, in Table~\ref{tab:stats}, we can observe that the overlap between adjacent timestamps, calculated via the Jaccard similarity index between subsequent node sets $V_t$ and $V_{t+1}$, is large and coherent over time, stabilizing on 64\% on average. Overall, the studied part of this platform emerges as a locally dense online community with temporally stable structural patterns and significant user engagement, which makes it interesting to study with our approach.


\subsection{Identification and Analysis of Archetypes}

We recall that a core aspect of our approach relies on identifying archetypes, namely typical patterns that can give us approximate descriptions of individual element behaviors. As we also noted in~\nameref{sec:archetype} section, archetypes can be easily defined for nodes and hyperedges. In this section, we aim to characterize user behaviors. Therefore, henceforth we refer to archetypes as related to nodes/users.

To define user archetypes, we combine platform-specific and linguistic features. Specifically, we outline eight archetypes based on these features and their values. The features we decide to use are, namely, \textit{(i)} score, \textit{(ii)} sentiment, and \textit{(iii)} toxicity. As far as the score is regarded, we compute the average post score for each user, then we normalize values in $[0,1]$. For the sentiment feature, we use the well-known VADER algorithm~\cite{HuEr14}, a rule-based sentiment classifier that assigns each text positive, negative, neutral, and compound sentiment intensity values. Out of these, the compound sentiment is a score in $[-1, 1]$ going from the most extreme negative ($-1$) to the most extreme positive ($1$) sentiment. For our purposes, we compute the average compound scores for each user and rescale values in $[0, 1]$. Moreover, for the toxicity feature, we compute its values with Detoxify~\cite{Detoxify}, a neural toxicity detection model that returns the probability that a text contains toxic language, and also in this case, we normalize the obtained values. Finally, to discern between high and low values for each feature, we set a threshold of 0.5, such that when a feature has a value higher than this threshold, it is considered high, and vice-versa.

The list of the eight archetypes defined is depicted in Table~\ref{table:reddit_archetypes}. Here, we can observe how setting the various combinations of the state of a feature (high or low) creates different archetypes. We defined these archetypes in order to encapsulate the inherent semantics of the discussions taking place in the studied dataset. For instance, when we refer to the \textit{HHL} combination, this represents users showing (H)igh scores, (H)igh sentiment (thus, in the range of positive levels), and (L)ow levels of toxicity. Hence, they could be interpreted as users who consistently contribute positively to the discussion by maintaining a positive and supportive attitude without engaging in toxic behavior. A detailed characterization of these archetypes is given in the following section.

\begin{table}[!ht]
\centering
\caption{User Archetypes Based on Score, Sentiment, and Toxicity.\label{table:reddit_archetypes}}
    \begin{tabular}{c|c|c|c}
        \toprule
        Score & Sentiment & Toxicity & \# \\
        \midrule
        H & H & L & 419 \\
        H & H & H & 21 \\
        H & L & L & 15286 \\
        H & L & H & 807 \\
        L & H & L & 267 \\ 
        L & H & H & 25 \\
        L & L & L & 3643 \\
        L & L & H & 469 \\
        \bottomrule
    \end{tabular}
\end{table}

\subsubsection{Archetype Characterization}

Having defined the archetypes, we are now able to provide a characterization of them, based on multiple psychological and social dimensions. More in detail, we describe the behaviors of these types with respect to the different aspects highlighted in the~\ref{sec:archetype-characterization} section. We recall that these aspects are, namely, the \textit{(i)} environment perception, the \textit{(ii)} emotions, and \textit{(iii)} morality. 

To control for class size imbalances, we compare average values for multiple features across the top 10 most archetypal users for each archetype, i.e., the most representative users for each category. To do so, we rank users according to their typicality, which we define as follows:
\begin{equation}
    Typicality(u) = \prod_f \alpha_{f} f(u)
\end{equation}

\begin{figure}
    \centering
    \includegraphics[width=.9\linewidth]{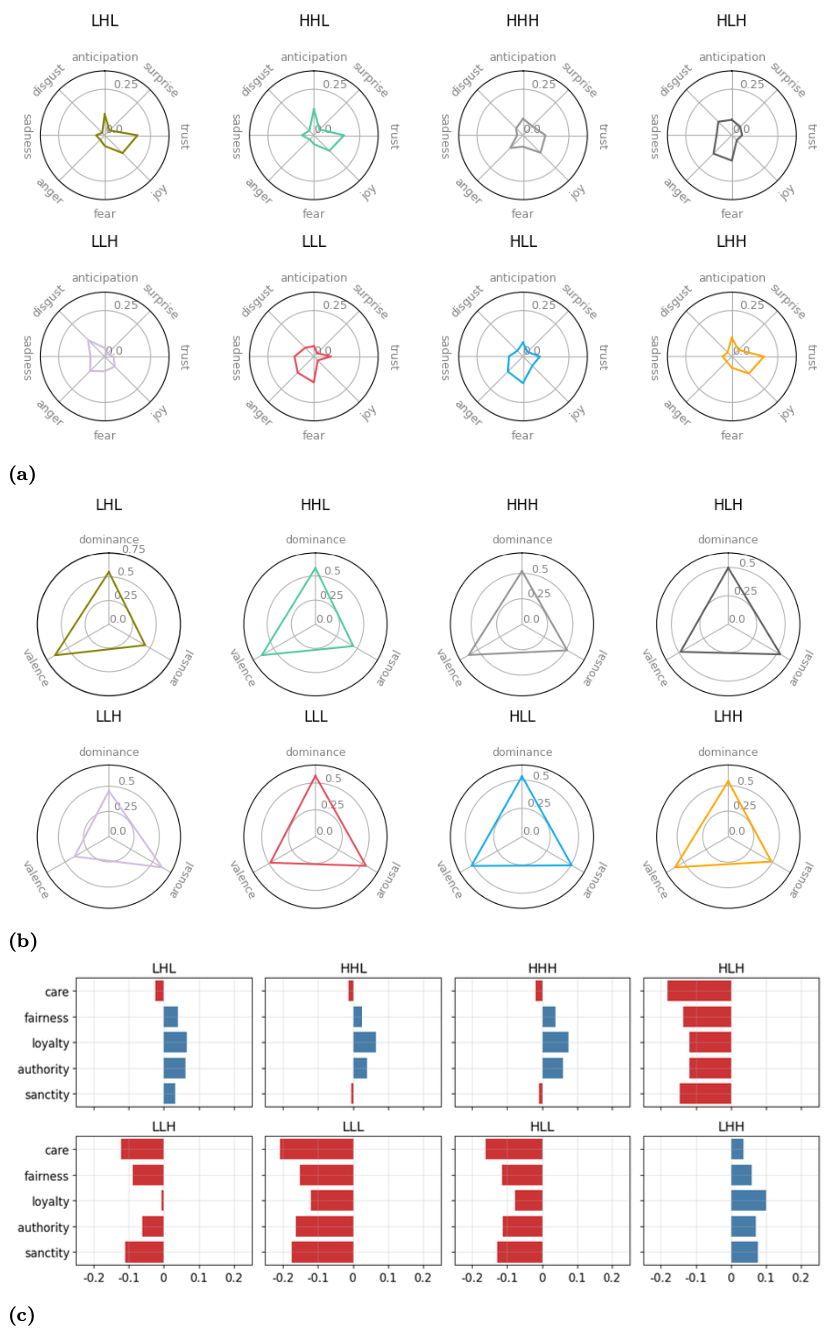}
    \caption{Archetype Profiles according to (a) Plutchik's wheel of emotions, (b) the Pleasure/Arousal/Dominance model and (c) Moral Foundations Theory.\label{fig:archetype-profile}}
\end{figure}

\noindent Here, $u$ is the target user; $f$ is a function computing $u$'s characteristic score, sentiment, or toxicity; $alpha_f$ is a coefficient that equals 1 if $u$'s archetype has a high value for $f$ and, conversely, -1 if $u$'s archetype has a low value for $f$. As an example, a \textit{HHL}'s typicality is maximized when its score is 1 (H), its sentiment value is 1 (H), and its toxicity value is 0 (L).
Along with the characterization, in Fig.~\ref{fig:archetype-profile}, we depict the profiles of each archetype. The top part of the figure represents the profiles according to Plutchik's wheel of emotions. The middle part, instead, represents the profiles according to the Pleasure/Arousal/Dominance model. Finally, the bottom part depicts the profiles according to the Moral Foundations theory.

\paragraph*{\textbf{LHL.}} With respect to their psychological features, \textit{LHL} users are characterized by high trust (0.18), joy (0.14), and anticipation (0.11), low arousal (0.44) and high valence (0.66). 
Hence, from their feature values, a possible interpretation can suggest them as users focusing on creating a supportive environment and emphasizing social cohesion, by manifesting a positive and optimistic outlook.
It is important to note that, for this archetype as well as all the following ones, the descriptors above can lead only to qualitative suggestions and interpretations of users' behaviors.

\paragraph*{\textbf{HHL.}} Such users are characterized by high trust (0.17), joy (0.12), and anticipation (0.14), low arousal (0.45) and high valence (0.64). 
Their psychological profile shows a generally positive outlook. 
From a moral focus perspective, the values suggest they mostly emphasize loyalty and, to an extent, authority.
We may hypothesize that these users would provide support in more prominent and/or critical situations, gaining greater acknowledgement from the community. 

\paragraph*{\textbf{HHH.}} With respect to their psychological features, \textit{HHH} users show signs of joy (0.13), trust (0.12), and anger (0.10), as well as high valence (0.60). 
This indicates that while they have a positive emotional tone and are capable of experiencing happiness and trust, they can also exhibit anger.
This leads to an interpretation of this archetype as an emotionally charged one, with a tendency to express both positive and negative tones.
Moreover, they mostly emphasize loyalty and authority moral dimensions.

\paragraph*{\textbf{HLH.}} This archetype shows high signs of anger (0.14), fear (0.14), and disgust (0.10).
Arousal (0.53) and Dominance (0.50) are average, while Valence is slightly below average (0.48).
This complex psychological profile hints at expressions of negative tones, perhaps toxic interactions.
High levels of fear could indicate a tendency to anticipate negative outcomes, which can make their discourse more alarmist. 
Disgust may contribute to a critical and often harsh tone. 
Average arousal and dominance could suggest a moderate level of engagement with the environment, whereas their slightly below-average valence could be a sign of negative interactivity.
Moral dimensions of care, fairness, loyalty, authority, and sanctity display negative values.
This moral outlook, combined with a negative emotional profile, contributes to an intuition of a provocative, even controversial, behavior.

\paragraph*{\textbf{LLH.}}
This archetype is characterized by high values for disgust (0.13), and anger (0.11), paired with below-average values for Valence (0.39) and Dominance (0.45), and high Arousal (0.61).
These emotional values manifest a full negative profile.
The archetype shows high arousal as well as strong emotional responses and engagement, but also low valence and dominance.
This could mean the users belonging to it often display dissatisfaction, leading to discontented interactions. Morally, \textit{LLH} users use vice-oriented language, focusing on sanctity/degradation and care/harm.
Low values in these areas could indicate strong criticism and a tendency to vent.

\paragraph*{\textbf{LLL.}}
This psychological profile outlines an emotional landscape characterized by high levels of sadness (0.11), fear (0.14), and anger (0.12). 
These predominant features suggest unhappiness with their environment.
Anger, although present, paired with low toxicity values, could suggest a profile expressing criticism in a non-confrontational way rather than being directly opposed.
With average levels of Valence, Arousal, and Dominance, their emotional responses are steady and moderated. 
From a moral perspective, \textit{LLL}'s language frequently involves vice-oriented terms, particularly focusing on sanctity/degradation, care/harm, and loyalty/betrayal. 
Their emphasis on sanctity could interestingly suggest a concern for moral purity, whereas the focus on care/harm could highlight preoccupation with injustice.
Their attention to loyalty/betrayal could indicate a sensitivity to perceived trust, which might align with their cautious yet critical nature. 

\paragraph*{\textbf{HLL}}
This psychological profile mostly elicits sadness (0.8), fear (0.14), and anger (0.12). 
Sadness and fear highlight anticipation of negative outcomes.
Anger drives the behavior of these users, moderated by average levels of valence, arousal, and dominance.

\paragraph*{\textbf{LHH.}} The users in this archetype are characterized by high levels of trust (0.18) and joy (0.13). They show average Dominance (0.54) and Arousal (0.49) values, as well as high Valence (.60).
Their high trust and joy could suggest they generally expect positive interactions from others and have a positive disposition.
The high Valence reflects an overall positive emotional tone. However, despite these positive traits, their high toxicity scores reveal a tendency to a duality, that we could interpret as a confrontational/provocative communication style. 
This dichotomy of positivity and toxicity pictures a complex profile where users' intentions may be good, but their execution can be damaging to others.

\subsubsection{Archetypes Transitions}
\label{sub:archetypes-transitions}

In this section, we aim to explore the temporal dimension of user interactions within the Scored.co community and understand whether significant transitions exist between the identified archetypes. This analysis not only helps in understanding the fluidity of user roles but also potentially provides valuable information for targeted interventions and community management.

In order to carry out this analysis, we propose the following methodology. Given the imbalanced class sizes (see Table~\ref{table:reddit_archetypes}) throughout the observation periods, we employ a null model to test the transition probabilities between archetypes. This null model computes the expected transition probabilities on a copy of our dataset where archetype labels are shuffled. The detailed steps are as follows:

\begin{enumerate}
    \item We generate $N = 500$ shuffled copies of the dataset, randomizing the archetype labels while preserving the overall distribution of interactions and activities;
    \item For each archetype pair $(A,B)$ in the shuffled copies we compute the transition probability $P(B|A)$, which represents the likelihood of a user transitioning from archetype $A$ to archetype $B$;
    \item We calculate the mean and standard deviation of $P(B|A)$ across all shuffled copies. These statistics are used to compute $z$-scores and $p$-values for the observed transition probabilities in the original dataset;
    \item Finally, transitions with $p$-values less than $0.01$ are considered statistically significant, indicating that these transitions considerably deviate from what is expected from the null model.
\end{enumerate}

\begin{figure}
    \centering
    \includegraphics[width=0.9\linewidth]{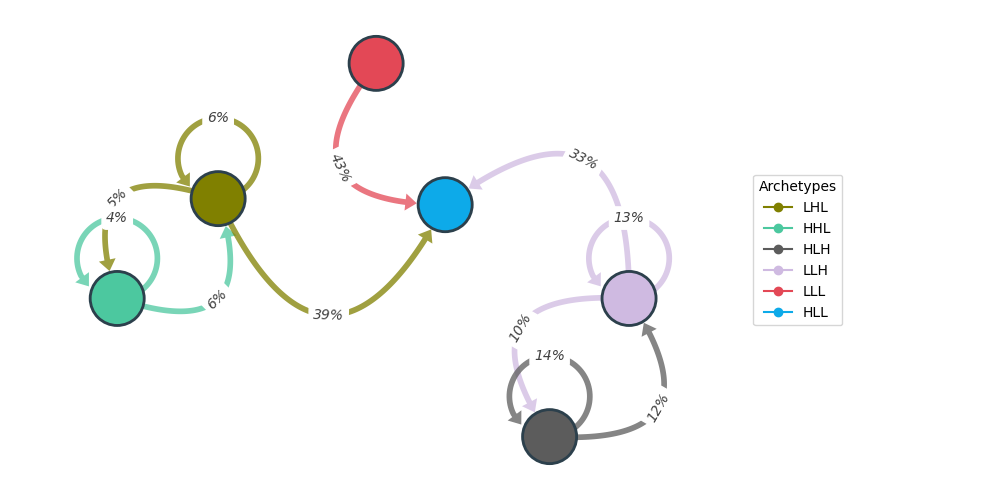}
    \caption{Transition Probabilities across archetypes (expressed in percentages). 
    Only statistically significant ones are shown ($p < 0.01$).}
    \label{fig:transitions}
\end{figure}

Fig.~\ref{fig:transitions} depicts the observed statistically significant transitions.
One notable transition is from LHL to HLL, occurring at a significant rate of 39.23\%.
The former ones are characterized by positive sentiments and low post scores, typically engaging in supportive and encouraging behaviors. This transition can indicate that even users who start with low recognition can, through consistent positive contributions, build credibility and eventually gain influence within the community.
HHL, another archetype defined by high trust, joy, and engagement, show a tendency to transition to LHL or, more rarely, to LHH. The slight shift to LHL suggests that HHL, despite their initially high engagement and recognition, may sometimes experience a reduction in visibility or engagement, leading them to adopt a more low-profile but still positive role. Also, we observe an occasional drift towards more toxic behaviors.
HLH, which have high post scores and negative sentiment, exhibit a significant probability of remaining in their "contentious" roles (13.71\%). However, there is also an 11.71\% likelihood of transitioning to LLH. This shift reflects a potential decrease in engagement and influence, where high-profile negative behavior can lead to more discontent and a less active state.
This also suggests a trajectory where persistent negativity can diminish a user's central role within the community.
LLH, who are driven by negative emotions and low post scores, present an interesting case with a substantial 33.29\% chance of shifting into HLL. This transition implies that their critical nature can eventually earn them credibility, provided they moderate their toxic behaviors. Another significant transition is observed from LLL to HLLL, with a remarkably high probability of 43.29\%. 

\subsubsection{Temporal Analysis of Archetype's Higher-order Interactions}

Above, we proposed a thorough characterization of a series of archetypes. At this point, an interesting aspect to study is the dynamics of archetypes within the social platform, as well as their evolution in time. To do so, we now focus on temporal trends of archetypes by examining some features on a monthly basis. We focus on two key metrics, that is, the average hyperdegree $\overline{hdeg}$ and the average degree $\overline{deg}$ of all nodes represented by each archetype.

\begin{figure}
    \centering
    \includegraphics[width=0.9\linewidth]{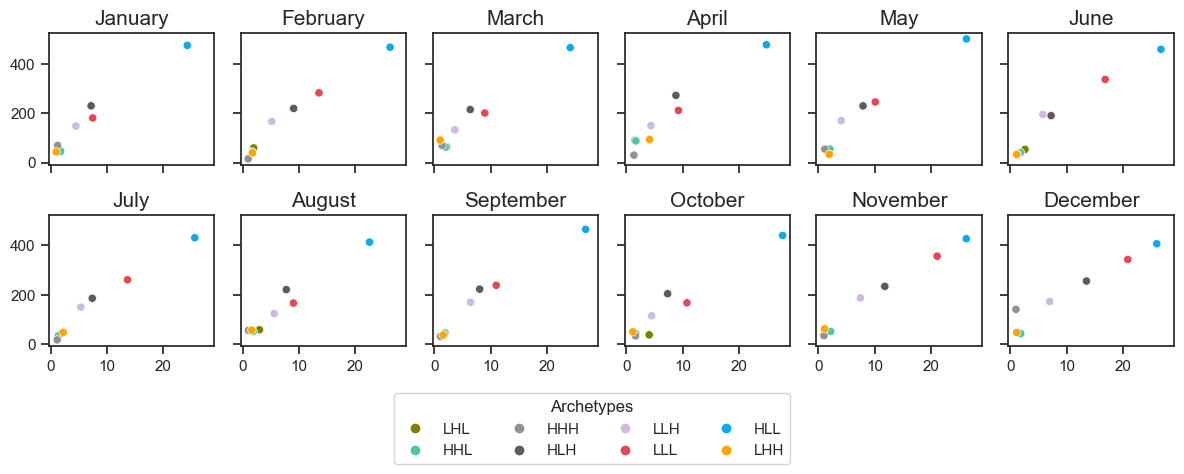}
    \caption{Monthly average hyperdegree (x-axis) and number of neighbors (y-axis) for each archetype.}
    \label{fig:monthly-archetypes}
\end{figure}

We consider the archetypes defined in the previous section, and we analyze the nodes each archetype represents in each month. More formally, for each archetype and for each node $u$, we compute its hyperdegree $hdeg(u)$ and its degree $deg(u)$, and we do this for each hypergraph ${\cal H}_t$, $t = 1,\dots,12$. In Fig.~\ref{fig:monthly-archetypes} we show the obtained results. The figure consists of 12 subplots, one for each hypergraph, corresponding to a different month. For each subplot, the $x$-axis represents the average hyperdegree $\overline{hdeg}$ of each archetype, while the $y$-axis represents the average degree $\overline{deg}$ of each archetype. From the analysis of this figure, different insights can be derived. First off, we can observe how the HLL archetype has a higher hyperdegree and degree than all other archetypes. This is somewhat expected as their characterization encompasses a considerable number of nodes, as we have seen in the~\nameref{sec:archetype-characterization} section. It is straightforward to observe how there are pairs of archetypes that are often collocated together w.r.t. these dimensions. This is the case of HLH and LLL, which appear very close in almost all monthly snapshots, although they are very different in size. The relatively small number of archetypes such as HHH and LHH is also reflected in their average hyperdegree and degree values: indeed, they remain stable during the months and have little to no fluctuations. This is somewhat expected, considering that these archetypes are characterized by emotionally charged and often provocative interactions, leading to fluctuating levels of activity and influence. This variability suggests that their presence is more dynamic, potentially driven by specific events or contentious discussions that peak at different times. Also in this case, these results could be somewhat enhanced by a more qualitative-oriented approach: indeed, aspects such as the content shared between archetypes would be of great importance in a more in-depth analysis of their interaction. Nevertheless, we believe the provided observation already indicates different implications. For instance, archetypes with stable engagement patterns are likely to be reliable contributors, therefore they could be taken into account in systematic operations of the platforms, such as moderation activities. Also, archetypes that show fluctuating feature values might be highly responsive to specific events or topics, which could fuel critical discussions. 

\subsection{Characterization of Discussions}

We recall that one of the objectives of our approach is to analyze higher-order entities. To do so, it offers the so-called hyperedge characterization function $\omega$, which we presented in the~\nameref{sub:hyperedge-characterization-function} section. We are now able to leverage it to describe the dataset with regard to some of the properties exhibited by the discussions.

First off, we focus on examining our dataset on a monthly basis, thus we consider each hypergraph ${\cal H}_t$, $t = 1, \dots, 12$. We do not focus on the overall hypergraph ${\cal H}$ due to the fact that we are interested in observing how the structure and the interactions within Scored.co evolve over time. Also, this approach allows for a more granular understanding of user behavior and interactions. For each hypergraph ${\cal H}_t$, we compute the \textit{s}-betweenness hyperedge centrality~\cite{aksoy2020hypernetwork}. This measure is a higher-order extension of classical graph betweenness centrality, and is computed on the line graph projection of the hypergraph to estimate the importance of hyperedges. We recall that the line graph of a hypergraph is a graph that represents the relationships between the hyperedges of the original hypergraph. In particular, there is a node in the line graph for each hyperedge in the hypergraph, and there is an edge between two nodes if the corresponding hyperedges share at least one node. After the computation of the centrality, we select the top 50 hyperedges in each hypergraph ${\cal H}_t$, and for each of we compute some descriptive features based on the linguistic productions of the contained nodes. In particular, we focus on the following features, namely, \textit{(i)} average word count, \textit{(ii)} average unique word count, and \textit{(iii)} average purity w.r.t. user archetypes. These values can be easily calculated via the $\omega$ function, specialized for each of them. For instance, the specialization for the average purity is given by $\omega_{\text{purity}}$ in the~\nameref{sub:hyperedge-characterization-function} section.

\begin{figure}
    \centering
    \includegraphics[width=\linewidth]{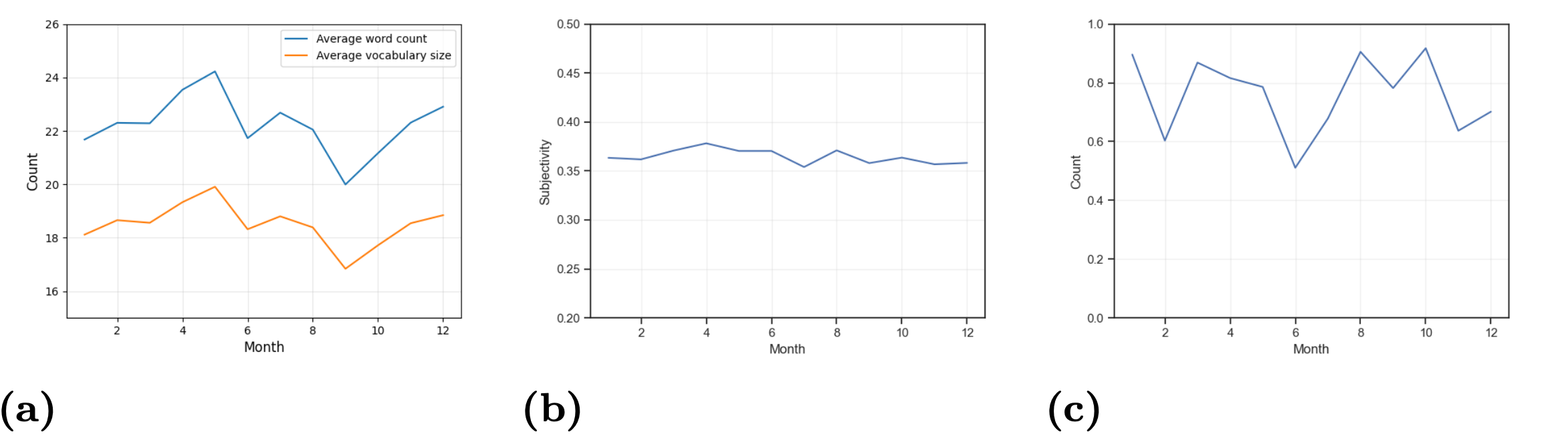}
    \caption{Average word count and average unique word count (a), average subjectivity (b), and average purity w.r.t. user archetypes (c) for the top 50 most central discussions.}
    \label{fig:disctrends}
\end{figure}

In Fig.~\ref{fig:disctrends}, we report the temporal trends of the average values of average word count, average unique word count, and average purity for the top 50 most prominent discussions. Note how the discussion here is another term for hyperedge: in fact, in our setting, a hyperedge represents a discussion between users in the social hypernetwork. From the analysis of this figure, we can derive different insights. First off, the average word count graph, in the left part of the figure, shows a relatively stable trend with slight fluctuations, indicating a consistent level of engagement in discussions over the year. There are some notable peaks, such as a slight increase around mid-year followed by a step-down and a subsequent rise towards the end of the year. Similarly, the average vocabulary size, which indicates the average unique word count, mirrors the previous measure but remains consistently lower, which is somewhat expected because of non-content words (e.g., articles, pronouns, etc.) appearing more frequently in texts. A slight increase in the middle of the year is also depicted here, suggesting periods of higher and lower lexical diversity in user interactions. The middle part of Fig.~\ref{fig:disctrends} reports the trend for average subjectivity. We recall that this metric evaluates how subjective (as opposed to objective) the discussions are. It remains relatively stable throughout the year, which indicates that discussions typically contain a balanced mix of personal opinions and factual statements. Indeed, this calls for a more qualitative-oriented analysis of these discussions, an approach often used in the content-based investigation of social platforms~\cite{CaKo23}. Finally, the right part of the image depicts the graph of the average purity w.r.t. the user archetypes. This refers to the consistency of user behavior and content relative to the defined archetypes. We can observe how this trend shows more variability compared to the previous features. There are sharp declines in some months, suggesting periods of more inconsistent behavior among users, while the different peaks could indicate times when user behavior is more aligned with archetypes. While all these insights are interesting, and show the applicability of our proposed approach, more in-depth analyses could be performed, as also noted above. Nevertheless, the various specializations of $\omega$ enable a series of different lenses through social interactions can be studied, especially in understudied platforms.

\section{Discussion and Implications}
\label{sec:discussion}

 In this study, we propose an approach for defining higher-order roles in a social hypernetwork and characterizing higher-order interaction. The approach culminates in a general framework allowing the definition of archetypes, i.e., high-level description of user roles, and the characterization of higher-order entities by taking into account their exhibited features and higher-order dynamics. Our experimental evaluation, conducted on the Scored.co platform, shows the advantages of the framework in enhancing user behavior modeling, mechanism of role evolution, and sentiment observability in a social platform characterized by higher-order interactions. We believe it is worth discussing the obtained results together with the framework's implications. To this end, we start by considering how our framework advances the task of modeling user behavior. We observed how the archetypes derived from our analysis can capture typical behavior patterns among users, revealing distinct user roles and interactions within hyperedges. For instance, we identified archetypes such as HHL (high score, high sentiment, low toxicity), which we can interpret as supportive users, and LLH (low score, low sentiment, high toxicity) which are often associated with disruptive behavior. These archetypes, along with their role-specific features, allow for interpretable modeling of user behaviors within certain contexts, as our analysis showed consistent participation patterns aligned with each archetype's characteristics. Note also that, while in our analysis we focus on three aspects only --- that is, score, sentiment, and toxicity --- ideally, one could consider more diverse aspects and also compare different combinations of them in order to have a multi-dimensional view of user roles. Moreover, it is worth noting how our analysis highlighted the temporal transitions between archetypes, thus revealing plausible mechanisms of role evolution. As we can observe in the~\nameref{sub:archetypes-transitions} section, a significant transition rate from the LHL archetype and the HLL one is present, and this shift could suggest a pathway where users with initially low visibility but positive engagement may, over time, gain influence and credibility within the community. Such an insight effectively highlights the practical value of our framework for tasks such as community management, where understanding role progression can help platforms foster constructive engagement. Finally, an important implication is the observation of sentiment and emotional dynamics. In fact, our framework integrates psychological and emotional dimensions, which in turn allows a heterogeneous archetype characterization, enhancing the analysis of sentiment trends within specific agglomerates of users. Indeed, we observed different insights that can be valuable in different tasks, such as designing community guidelines and identifying disruptive sentiment dynamics. For instance, we noted that LLL archetype users frequently displayed high levels of sadness, fear, and anger, with low toxicity scores, indicating that these users express negative emotions without leaning into harmful language, which could disrupt the experience of other users. Nevertheless, such insights could be of great use in contexts such as analyzing health discourses and information~\cite{ChSh14,Record*18}.

\section{Conclusion}
\label{sec:conclusion}

The identification of node roles within complex networks is significant when analyzing their dynamics and function. To advance in this context, in this paper we proposed a multi-dimensional, general framework to characterize nodes and hyperedges in a social hypernetwork. The aim of our framework is two-fold: i) to characterize nodes and hyperedges taking into account their exhibited features and higher-order dynamics, and ii) to define ``archetypes'', serving as a template to represent higher-order roles of nodes. 
Our framework consists of different components, namely, hyperedge ensembles, hyperedge- and node-based characterization functions. The combination of such components allows one to analyze the social hypernetwork from a more general point of view; each component can also be exploited in a standalone fashion, thus enabling a detailed analysis of the single aspect it captures. 
To assess the effectiveness of the framework, we carried out an exhaustive experimental campaign on Scored.co, an understudied social platform, focusing on different aspects such as the characterization of nodes as well as their behaviors and surroundings. 

This paper is a starting point for future research with various possible extensions.
First off, the framework can be applied \textit{as it is} on other (possibly understudied) social platforms such as Bluesky~\cite{FaRo24}.
Second, it would be interesting to compare archetypes based on the set of features selected for defining them. Indeed, an approach to identify the archetype best representing a particular set of nodes could be defined as an optimization problem, and meta-heuristics solutions could be employed to solve it~\cite{sohail2023genetic}. 
Another future direction would be studying role dynamics in crisis situations. Indeed, we could apply the framework to study social hypernetworks in crisis situations, e.g., natural disasters and pandemics, to understand how roles and behaviors shift under stress and how information and support are mobilized~\cite{BaCaTe21,Zhang*19}. 
Exploiting the temporal component, we would integrate this framework about higher-order roles in a multi-dimensional analysis of group evolution in temporal data \cite{failla2024describing}.
Finally, an interesting challenge would be the analysis of cross-platform role identification, and in particular the investigation of the consistency of identified roles across multiple social platforms. This could help in understanding whether certain archetypes are platform-specific or they can represent more universal roles in online social interactions.
\section*{Acknowledgments}

This work is supported by (i) the European Union – Horizon 2020 Program under the scheme “INFRAIA-01-2018-2019 – Integrating Activities for Advanced Communities”, Grant Agreement n.871042, ''SoBigData++: European Integrated Infrastructure for Social Mining and Big Data Analytics" (\url{http://www.sobigdata.eu}); (ii) SoBigData.it
which receives funding from the European Union – NextGenerationEU – National Recovery and Resilience Plan (Piano Nazionale di Ripresa e Resilienza, PNRR) – Project: ''SoBigData.it – Strengthening the Italian RI for Social Mining and Big Data Analytics" – Prot. IR0000013 – Avviso n. 3264 del 28/12/2021; (iii) EU NextGenerationEU programme under the funding schemes PNRR-PE-AI FAIR (Future Artificial Intelligence Research).

\bibliography{bibliography}

\begin{thebibliography}{10}

\bibitem{aksoy2020hypernetwork}
S.~G. Aksoy, C.~Joslyn, C.~O. Marrero, B.~Praggastis, and E.~Purvine.
\newblock Hypernetwork science via high-order hypergraph walks.
\newblock {\em EPJ Data Science}, 9(1):16, 2020.

\bibitem{AlGaradi*18}
M.~A. Al-Garadi, K.~D. Varathan, S.~D. Ravana, E.~Ahmed, G.~Mujtaba, M.~U.~S. Khan, and S.~U. Khan.
\newblock Analysis of online social network connections for identification of influential users: Survey and open research issues.
\newblock {\em ACM Computing Surveys (CSUR)}, 51(1):1--37, 2018.
\newblock ACM.

\bibitem{Bajardi*11}
P.~Bajardi, A.~Barrat, F.~Natale, L.~Savini, and V.~Colizza.
\newblock Dynamical patterns of cattle trade movements.
\newblock {\em PloS one}, 6(5):e19869, 2011.

\bibitem{BaCaTe21}
V.~Basile, F.~Cauteruccio, and G.~Terracina.
\newblock How dramatic events can affect emotionality in social posting: the impact of covid-19 on reddit.
\newblock {\em Future Internet}, 13(2):29, 2021.
\newblock MDPI.

\bibitem{Battiston*20}
F.~Battiston, G.~Cencetti, I.~Iacopini, V.~Latora, M.~Lucas, A.~Patania, J.-G. Young, and G.~Petri.
\newblock Networks beyond pairwise interactions: Structure and dynamics.
\newblock {\em Physics Reports}, 874:1--92, 2020.
\newblock Elsevier.

\bibitem{BhCoMu11}
S.~Bhagat, G.~Cormode, and S.~Muthukrishnan.
\newblock Node classification in social networks.
\newblock {\em Social network data analytics}, pages 115--148, 2011.

\bibitem{BoHoJo04}
P.~Bonacich, A.~C. Holdren, and M.~Johnston.
\newblock Hyper-edges and multidimensional centrality.
\newblock {\em Social networks}, 26(3):189--203, 2004.
\newblock Elsevier.

\bibitem{BrKr11}
R.~Brendel and H.~Krawczyk.
\newblock Primary role identification in dynamic social networks.
\newblock In {\em Proc.\ of 2011 International Conference on Computational Aspects of Social Networks (CASoN 2011)}, pages 54--59, 2011.
\newblock IEEE.

\bibitem{BuGo14}
C.~Buntain and J.~Golbeck.
\newblock Identifying social roles in reddit using network structure.
\newblock In {\em Proc.\ of the 23rd international conference on World Wide Web (WWW 2014)}, pages 615--620, 2014.
\newblock ACM.

\bibitem{Cauteruccio*22}
F.~Cauteruccio, E.~Corradini, G.~Terracina, D.~Ursino, and L.~Virgili.
\newblock Investigating reddit to detect subreddit and author stereotypes and to evaluate author assortativity.
\newblock {\em Journal of Information Science}, 48(6):783--810, 2022.
\newblock SAGE.

\bibitem{CaKo23}
F.~Cauteruccio and Y.~Kou.
\newblock Investigating the emotional experiences in esports spectatorship: The case of league of legends.
\newblock {\em Information Processing \& Management}, 60(6):103516, 2023.

\bibitem{cima2024great}
L.~Cima, A.~Trujillo, M.~Avvenuti, and S.~Cresci.
\newblock The great ban: Efficacy and unintended consequences of a massive deplatforming operation on reddit.
\newblock In {\em Companion Publication of the 16th ACM Web Science Conference}, pages 85--93, 2024.

\bibitem{ChSh14}
M.~De~Choudhury and S.~De.
\newblock Mental health discourse on reddit: Self-disclosure, social support, and anonymity.
\newblock In {\em Proceedings of the international AAAI conference on web and social media}, volume~8, pages 71--80, 2014.

\bibitem{Dehghan*23}
A.~Dehghan, K.~Siuta, A.~Skorupka, A.~Dubey, A.~Betlen, D.~Miller, W.~Xu, B.~Kami{\'n}ski, and P.~Pra{\l}at.
\newblock Detecting bots in social-networks using node and structural embeddings.
\newblock {\em Journal of Big Data}, 10(1):119, 2023.
\newblock Springer.

\bibitem{failla2024describing}
A.~Failla, R.~Cazabet, G.~Rossetti, and S.~Citraro.
\newblock Describing group evolution in temporal data using multi-faceted events.
\newblock {\em Machine Learning}, 113(10):7591--7615, 2024.

\bibitem{failla2023attributed}
A.~Failla, S.~Citraro, and G.~Rossetti.
\newblock Attributed stream hypergraphs: temporal modeling of node-attributed high-order interactions.
\newblock {\em Applied Network Science}, 8(1):31, 2023.

\bibitem{scoredHD}
A.~Failla, S.~Citraro, G.~Rossetti, and F.~Cauteruccio.
\newblock Scored.co hypernetwork dataset, 2024.

\bibitem{FaRo24}
A.~Failla and G.~Rossetti.
\newblock “i’m in the bluesky tonight”: Insights from a year worth of social data.
\newblock {\em PloS one}, 19(11):e0310330, 2024.

\bibitem{HaRi21}
J.~Hacker and K.~Riemer.
\newblock Identification of user roles in enterprise social networks: method development and application.
\newblock {\em Business \& Information Systems Engineering}, 63(4):367--387, 2021.
\newblock Springer.

\bibitem{haidt2004intuitive}
J.~Haidt and C.~Joseph.
\newblock Intuitive ethics: How innately prepared intuitions generate culturally variable virtues.
\newblock {\em Daedalus}, 133(4):55--66, 2004.

\bibitem{Detoxify}
L.~Hanu and {Unitary team}.
\newblock Detoxify.
\newblock Github. https://github.com/unitaryai/detoxify, 2020.

\bibitem{hopp2021extended}
F.~R. Hopp, J.~T. Fisher, D.~Cornell, R.~Huskey, and R.~Weber.
\newblock The extended moral foundations dictionary (emfd): Development and applications of a crowd-sourced approach to extracting moral intuitions from text.
\newblock {\em Behavior research methods}, 53:232--246, 2021.

\bibitem{Huang*14}
S.~Huang, T.~Lv, X.~Zhang, Y.~Yang, W.~Zheng, and C.~Wen.
\newblock Identifying node role in social network based on multiple indicators.
\newblock {\em PloS one}, 9(8):e103733, 2014.
\newblock Public Library of Science San Francisco.

\bibitem{HuEr14}
C.~Hutto and E.~Gilbert.
\newblock Vader: A parsimonious rule-based model for sentiment analysis of social media text.
\newblock In {\em Proc.\ of the International AAAI Conference on Web and Social Media (ICWSM 2014)}, volume~8, pages 216--225, 2014.

\bibitem{Junchen*21}
J.~Jin, M.~Heimann, D.~Jin, and D.~Koutra.
\newblock Toward understanding and evaluating structural node embeddings.
\newblock {\em ACM Transactions on Knowledge Discovery from Data (TKDD)}, 16(3):1--32, 2021.
\newblock ACM.

\bibitem{Kou*18}
Y.~Kou, C.~M. Gray, A.~L. Toombs, and R.~S. Adams.
\newblock Understanding social roles in an online community of volatile practice: A study of user experience practitioners on reddit.
\newblock {\em ACM Transactions on Social Computing}, 1(4):1--22, 2018.
\newblock ACM.

\bibitem{lazer2009computational}
D.~Lazer, A.~Pentland, L.~Adamic, S.~Aral, A.-L. Barab{\'a}si, D.~Brewer, N.~Christakis, N.~Contractor, J.~Fowler, M.~Gutmann, et~al.
\newblock Computational social science.
\newblock {\em Science}, 323(5915):721--723, 2009.

\bibitem{mehrabian1974approach}
A.~Mehrabian and J.~A. Russell.
\newblock {\em An approach to environmental psychology.}
\newblock the MIT Press, 1974.

\bibitem{MeFaBa24}
A.~Mekacher, M.~Falkenberg, and A.~Baronchelli.
\newblock The koo dataset: An indian microblogging platform with global ambitions.
\newblock In {\em Proceedings of the International AAAI Conference on Web and Social Media}, volume~18, pages 1991--2002, 2024.

\bibitem{mekacher2022can}
A.~Mekacher and A.~Papasavva.
\newblock " i can’t keep it up." a dataset from the defunct voat. co news aggregator.
\newblock In {\em Proceedings of the International AAAI Conference on Web and Social Media}, volume~16, pages 1302--1311, 2022.

\bibitem{mohammad2018obtaining}
S.~Mohammad.
\newblock Obtaining reliable human ratings of valence, arousal, and dominance for 20,000 english words.
\newblock In {\em Proceedings of the 56th annual meeting of the association for computational linguistics (volume 1: Long papers)}, pages 174--184, 2018.

\bibitem{mohammad2013crowdsourcing}
S.~M. Mohammad and P.~D. Turney.
\newblock Crowdsourcing a word--emotion association lexicon.
\newblock {\em Computational intelligence}, 29(3):436--465, 2013.

\bibitem{PaPeVa17}
A.~Patania, G.~Petri, and F.~Vaccarino.
\newblock The shape of collaborations.
\newblock {\em EPJ Data Science}, 6:1--16, 2017.
\newblock Springer.

\bibitem{patania2017shape}
A.~Patania, G.~Petri, and F.~Vaccarino.
\newblock The shape of collaborations.
\newblock {\em EPJ Data Science}, 6:1--16, 2017.

\bibitem{Patel*24}
J.~Patel, P.~Paudel, E.~De~Cristofaro, G.~Stringhini, and J.~Blackburn.
\newblock idrama-scored-2024: A dataset of the scored social media platform from 2020 to 2023.
\newblock In {\em Proceedings of the International AAAI Conference on Web and Social Media}, volume~18, pages 2014--2024, 2024.

\bibitem{plutchik1980general}
R.~Plutchik.
\newblock A general psychoevolutionary theory of emotion.
\newblock In {\em Theories of emotion}, pages 3--33. Elsevier, 1980.

\bibitem{QuBo24}
D.~Quelle and A.~Bovet.
\newblock Bluesky: Network topology, polarisation, and algorithmic curation.
\newblock {\em arXiv preprint arXiv:2405.17571}, 2024.

\bibitem{Record*18}
R.~A. Record, W.~R. Silberman, J.~E. Santiago, and T.~Ham.
\newblock I sought it, i reddit: Examining health information engagement behaviors among reddit users.
\newblock {\em Journal of health communication}, 23(5):470--476, 2018.

\bibitem{ReZaSo19}
R.~Recuero, G.~Zago, and F.~Soares.
\newblock Using social network analysis and social capital to identify user roles on polarized political conversations on twitter.
\newblock {\em Social media+ society}, 5(2):2056305119848745, 2019.
\newblock SAGE.

\bibitem{RoAlSa21}
B.~Rozemberczki, C.~Allen, and R.~Sarkar.
\newblock Multi-scale attributed node embedding.
\newblock {\em Journal of Complex Networks}, 9(2):cnab014, 2021.
\newblock Oxford University Press.

\bibitem{SeStLe16}
V.~Sekara, A.~Stopczynski, and S.~Lehmann.
\newblock Fundamental structures of dynamic social networks.
\newblock {\em Proceedings of the national academy of sciences}, 113(36):9977--9982, 2016.
\newblock National Academy of Sciences.

\bibitem{Singhal01}
A.~Singhal.
\newblock {Modern information retrieval: A brief overview}.
\newblock {\em IEEE Data Engineering Bullettin}, 24(4):35--43, 2001.

\bibitem{sohail2023genetic}
A.~Sohail.
\newblock Genetic algorithms in the fields of artificial intelligence and data sciences.
\newblock {\em Annals of Data Science}, 10(4):1007--1018, 2023.

\bibitem{Temdee*06}
P.~Temdee, B.~Thipakorn, B.~Sirinaovakul, and H.~Schelhowe.
\newblock Of collaborative learning team: An approach for emergent leadership roles identification by using social network analysis.
\newblock In {\em Proc.\ of Technologies for E-Learning and Digital Entertainment: First International Conference}, pages 745--754, 2006.
\newblock Springer.

\bibitem{torres2021and}
L.~Torres, A.~S. Blevins, D.~Bassett, and T.~Eliassi-Rad.
\newblock The why, how, and when of representations for complex systems.
\newblock {\em SIAM Review}, 63(3):435--485, 2021.

\bibitem{WaLuYu14}
P.~Wang, J.~L{\"u}, and X.~Yu.
\newblock Identification of important nodes in directed biological networks: A network motif approach.
\newblock {\em PloS one}, 9(8):e106132, 2014.

\bibitem{Zhang*19}
C.~Zhang, C.~Fan, W.~Yao, X.~Hu, and A.~Mostafavi.
\newblock Social media for intelligent public information and warning in disasters: An interdisciplinary review.
\newblock {\em International Journal of Information Management}, 49:190--207, 2019.

\bibitem{Zhou*22}
J.~Zhou, L.~Liu, W.~Wei, and J.~Fan.
\newblock Network representation learning: from preprocessing, feature extraction to node embedding.
\newblock {\em ACM Computing Surveys (CSUR)}, 55(2):1--35, 2022.

\bibitem{ZhWuJi19}
X.~Zhou, B.~Wu, and Q.~Jin.
\newblock User role identification based on social behavior and networking analysis for information dissemination.
\newblock {\em Future Generation Computer Systems}, 96:639--648, 2019.
\newblock Elsevier.

\end{thebibliography}

\end{document}